\documentclass[sigconf]{acmart}

\graphicspath{{Figures/}}
\usepackage{tabularray}
\usepackage{float}
\usepackage{graphicx}
\usepackage[export]{adjustbox}
\usepackage{caption}
\usepackage{subcaption}
\usepackage{todonotes}
\usepackage{wrapfig}
\usepackage{array}
\usepackage{placeins}
\usepackage{multicol}
\usepackage{xspace}
\usepackage{listings}
\usepackage{tabularx}
\usepackage{booktabs}

\copyrightyear{2026}
\acmYear{2026}
\setcopyright{cc}
\setcctype{by-nc-nd}
\acmConference[C\&C '26]{Creativity and Cognition}{July 13--16, 2026}{London, United Kingdom}
\acmBooktitle{Creativity and Cognition (C\&C '26), July 13--16, 2026, London, United Kingdom}
\acmDOI{10.1145/3803784.3816859}
\acmISBN{979-8-4007-2583-8/2026/07}

\newcommand{\MM}{\emph{Memetic Mixer}\xspace}

\begin{document}

\title{Language as a Material Interface for Creative LLM Interaction}

\author{Jon McCormack}
\orcid{0000-0001-6328-5064}
\affiliation{%
  \institution{SensiLab, Monash University}
  \city{Caulfield East}
  \state{Victoria}
  \country{Australia}}
\email{Jon.McCormack@monash.edu}
\author{Tace McNamara}
\orcid{0009-0009-7649-2420}
\affiliation{%
  \institution{SensiLab, Monash University}
  \city{Caulfield East}
  \state{Victoria}
  \country{Australia}}
\email{Tace.McNamara@monash.edu}
\author{Chen Wang}
\orcid{0009-0008-0597-1064}
\affiliation{%
  \institution{SensiLab, Monash University}
  \city{Caulfield East}
  \state{Victoria}
  \country{Australia}}
\email{Chloe.Wang1@monash.edu}
\author{Maria Teresa Llano}
\orcid{0000-0002-4898-1755}
\affiliation{%
  \institution{University of Sussex}
  \city{Brighton}
  \country{United Kingdom}}
\email{Teresa.Llano@sussex.ac.uk}

\renewcommand{\shortauthors}{McCormack et al.}

\begin{abstract}
Although directive prompting is the predominant way to interact with Large Language Models (LLMs), many creative practices rely on language that is open-ended, associative, phonaesthetic, symbolic, and that unfolds across multiple temporalities. In this work, we explore how creative practitioners might work with AI systems when language is treated not merely as instruction but as material. We conducted an ecological two-week study with four creative practitioners using a design probe: the \MM, a tangible interactive device that constrains interaction with an LLM. Analysis of post-study interviews and device logs identified distinct modes of material language use and temporalities that shaped each participant's engagement with AI and their creative practice. We reflect on these findings and contribute design considerations that support open-ended interaction with AI in creative practice.
\end{abstract}



\received{29 January 2026}


\begin{teaserfigure}
    \centering
    \includegraphics[width=\linewidth]{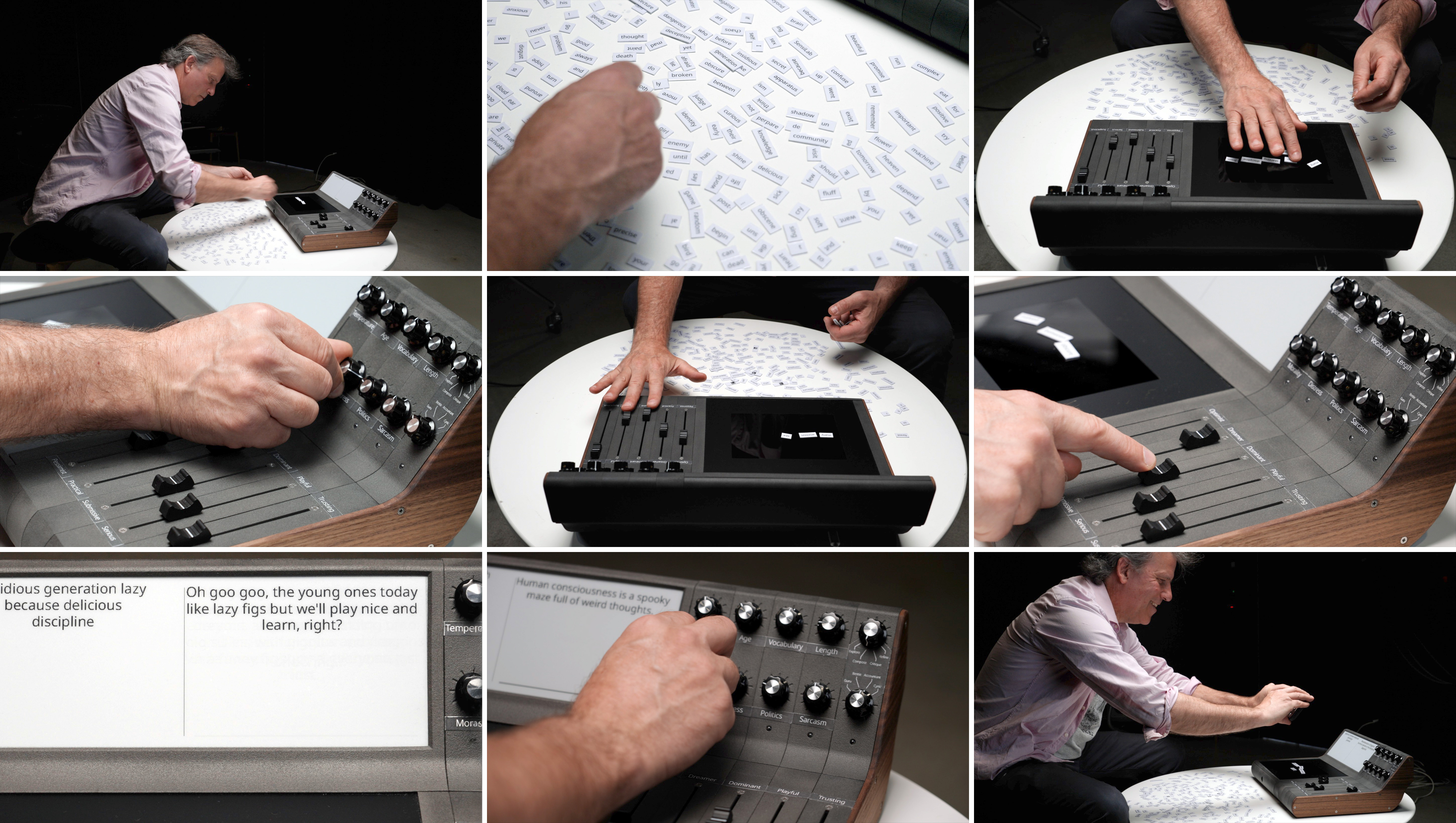}
    \caption{Hands-on interactions with the \MM}
    \label{fig:mm-interactions}
\end{teaserfigure}

\maketitle

\section{Introduction}
\label{sec:introduction}

Large language models (LLMs) are now commonplace tools used to find information, provide explanations, and generate creative artefacts. However, across these modes of usage, chat interfaces constrain LLM interactions to discrete exchanges of prompt and response. While literal or directed language -- such as giving unambiguous instructions -- is important in many contexts, language can also be used in open-ended and creative contexts. This necessitates considering language not only as a semantic interface, but foregrounding its \textit{materiality:} as mood and affect, giving-shape, sound and image, and physical form. 

Language as a material interface is open-ended and associative. It is used across continuous gradients in dialogue, and in the atemporal dimension of symbolism. Language becomes multisensory through its numerous physical forms; from the visual personality of a handwritten note, 
to the physical sensation when making utterances with our voices. These material aspects of language are unrecognised by LLMs, which treat text exclusively as digital tokens \cite{ali_tokenizer_2024}. To fulfil the richness of human language, materiality should be present in creative interactions with LLMs.

This study aims to better understand how language can be used as a material interface for creative LLM interaction. 
We recruited four creative practitioners to use a tangible AI device (the \MM \cite{mcnamaraMixer:2025}) into their creative practice to interact with an LLM as they saw fit, for a two-week period. 
As a \textit{design probe} 
the device was used to explore how an individual might \emph{want} to incorporate AI into their creative practice and 
how language and materiality shape both their interaction with LLMs and their creative practice.

Analysis of post-use interviews and device interaction logs revealed each of the four participants used a distinct mode of material language and a unique scale of temporality when reflecting on creative practice and AI interaction. The study evidences the specificity of individuals' creative practice and the possibilities and tensions that arise when interacting with AI, contributing:

\begin{itemize}
    \item Qualitative accounts of in-depth experience of creative practitioners interactions with AI.
    \item Examples of how language can be used as material interface for AI in creative practice.
\end{itemize}

%
\section{Background and Related Work}
\label{sec:background}

\subsection{Materiality of language}
\label{ss:materiality__of_language}
Scholarship has moved beyond treating language as a simple vehicle for transmitting ideas, understanding it instead as embodied, mediated, and embedded in objects and practices that produce tangible social and economic effects \cite{cavanaugh2017language,pennycook2024after}. This materiality extends to creative practice. \citet{jacucci2007performative} argue that material artefacts support collective creativity through their sensory and representational qualities, cautioning against a narrow focus on verbal expression alone. Despite its relevance to creative practice, the role of material language in interactions with LLMs remains under explored. 

\subsection{Temporality and Creativity}
Creativity and flow states \cite{alameda_brain_2022,mihaly_csikszentmihalyi_creativity_1996} involve a subjective experience of time where intense focus can restructure temporal experience \cite{van2016restructuring} and moments can feel accelerated, slowed, or suspended \cite{mainemelis2002time}. The creative process also unfolds non-linearly, future visions shape revision of creative works and guide present action \cite{palani2024evolving}. Iterative rather than linear approaches tend to support more original outcomes \cite{tolkamp2025creativity}, a pattern reflected in writing tools for LLM-supported creativity \cite{chakrabarty_art_2024,sood:ICCC25,carrera:2025,suh:CHI24}.

\subsection{Interfacing with AI} 

Most LLM writing tools rely on conversational prompting for interpretation, rewriting, and suggestion \cite{zhou:CC2024}, but research on non-instructional interfaces is emerging, where tangible materials serve as embodied inputs rather than direct controls. Examples include arranging paper story elements \cite{kuntong:TEI25}, clay-based character representations \cite{Hesselbarth:CC25}, and hand-shaped materials for prompting image or narrative generation \cite{balasubramani:2024,menon:2024mind}. Recent work has also used tangible controls to interface with AI such as \textit{GenFrame} \cite{kun:DIS24}, a physical frame with knobs for modifying AI-generated images, and the \MM \cite{mcnamaraMixer:2025} which uses knobs and faders inspired by audio mixing consoles and physical word tiles to facilitate embodied interaction with an LLM. However, interfaces which use language as a material for AI interaction and how this fits into creative practice remains under explored, a gap we address by using the \MM as a design probe (Section \ref{ss:theMM}).

\subsection{The \MM as Design Probe}
\label{ss:theMM}

\begin{figure*}
    \centering
    \includegraphics[width=\linewidth]{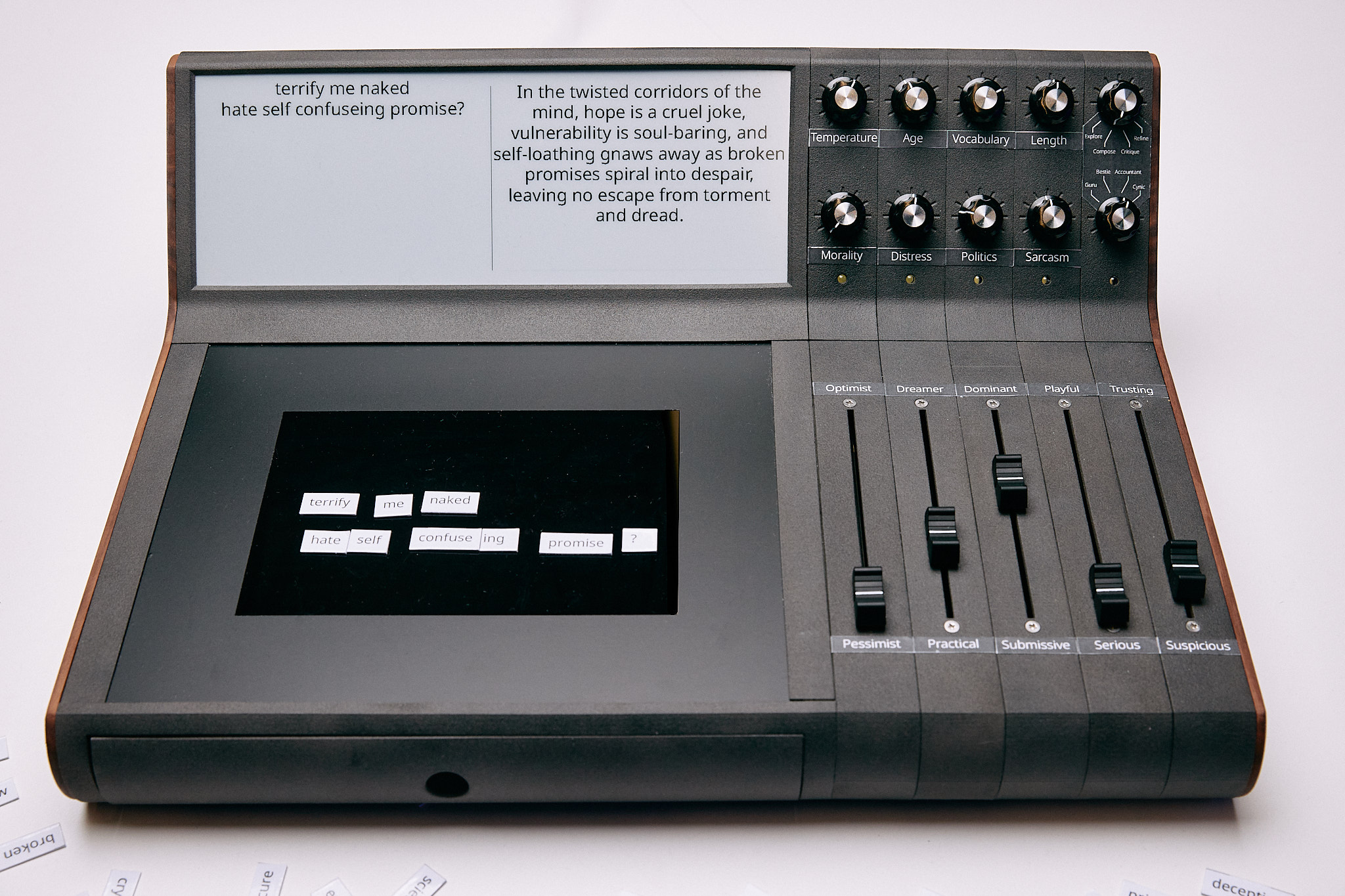}
    \caption{The \MM, used as a design probe with four creative practitioners. To interact with the device, word tiles are placed on the dark rectangular region and the e-screen at the top displays words as they as placed (left) and the device's response to them (right). The bottom right faders represent personality traits, while the knobs above provide different ways to direct the LLM.}
    \label{fig:mm}
\end{figure*}

The \MM (Figures \ref{fig:mm-interactions} and \ref{fig:mm}) affords various modes of interaction with an LLM that are not centred on the user providing direct prompt exchanges or instructions. We briefly outline the device's interface and controls below. For more comprehensive technical details on the device and its design, please refer to  \cite{mcnamaraMixer:2025,mccormack_mimetic_2024}.

\begin{figure}
     \centering
        \includegraphics[width=\linewidth]{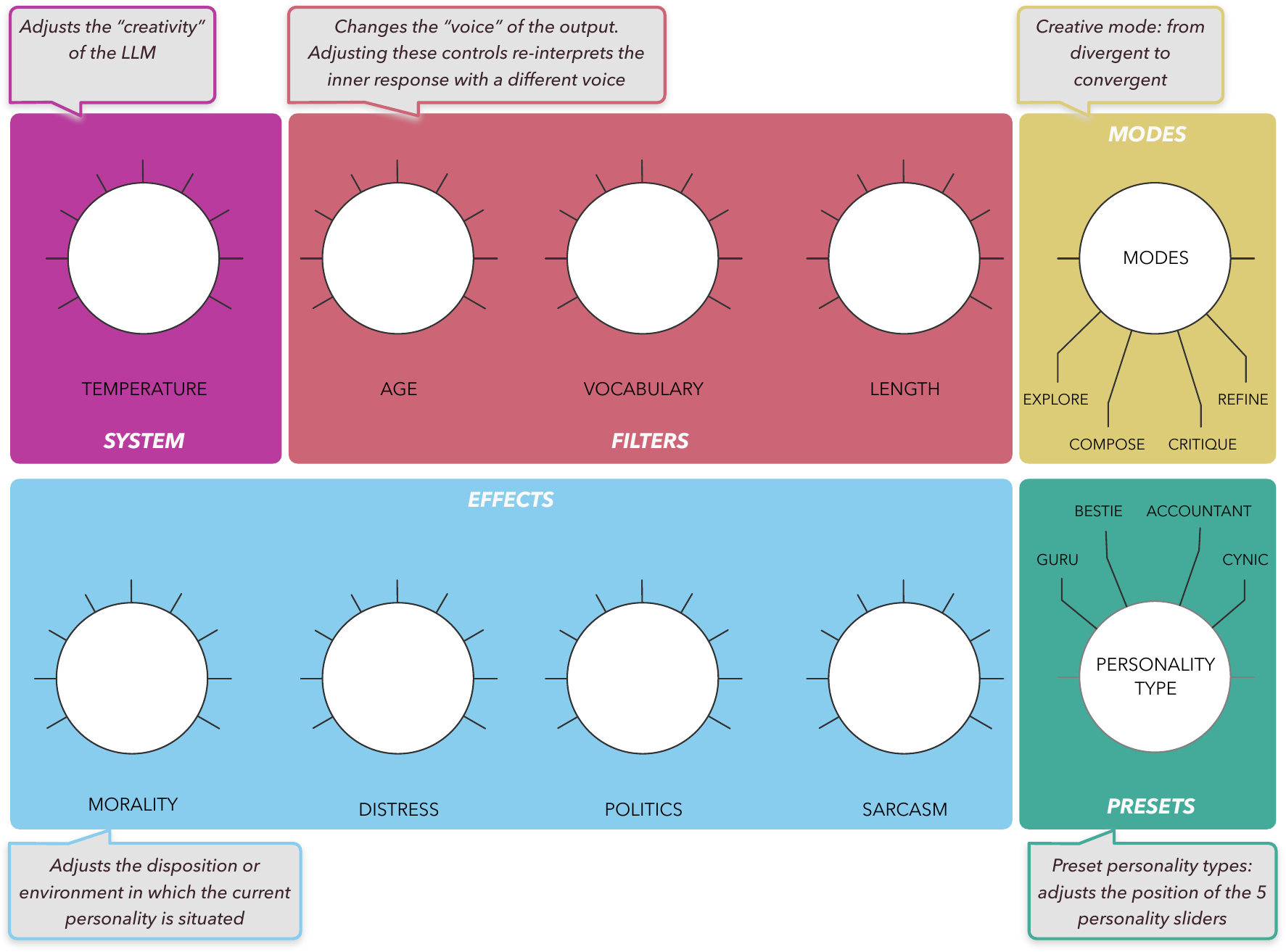} \\
        \includegraphics[width=0.9\linewidth]{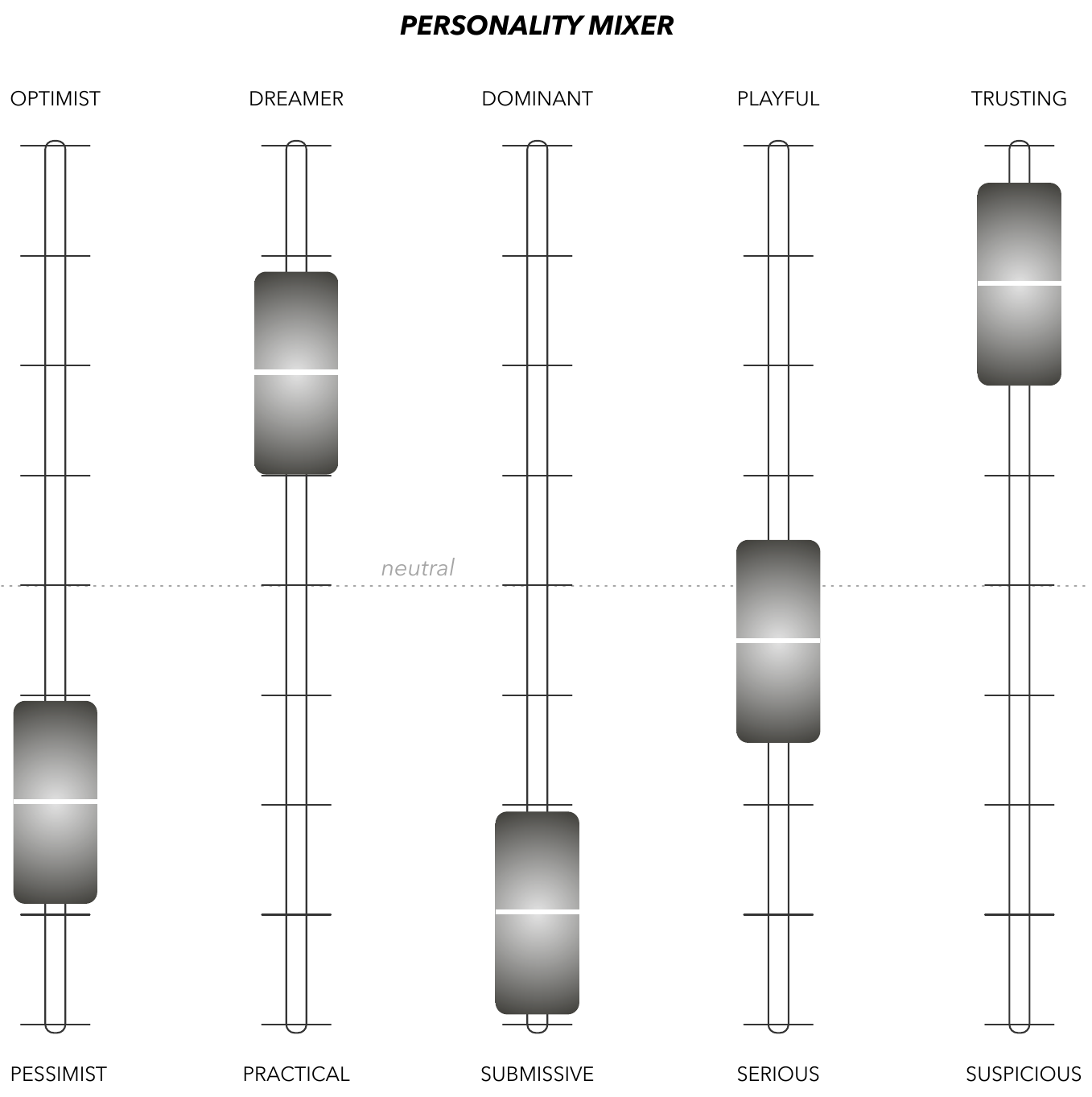}
     \caption{The \MM's control knobs (top) adjust the style, mood, length and mode of responses, while the personality mixer (bottom) uses faders to set the personality of the LLM responses.}
     \label{fig:knobsAndSliders}
\end{figure}

\subsubsection{Interaction through Tangible Language}
The device uses physical \emph{word tiles}, similar to those found in Magnetic Poetry \cite{kapell_magnetic_1997}, to provide input to the LLM. Users physically place words on the device, adjust the controls and a response is generated on the display (Figure \ref{fig:mm-interactions}).  A pre-defined collection of around 120 commonly-used English words and word-stems (e.g. ``ly'') are recognised by an internal machine vision system when placed on the rectangular input \textit{slate} on the front of the device (Figure \ref{fig:mm}).

\begin{figure*}
    \centering
    \includegraphics[width=\linewidth]{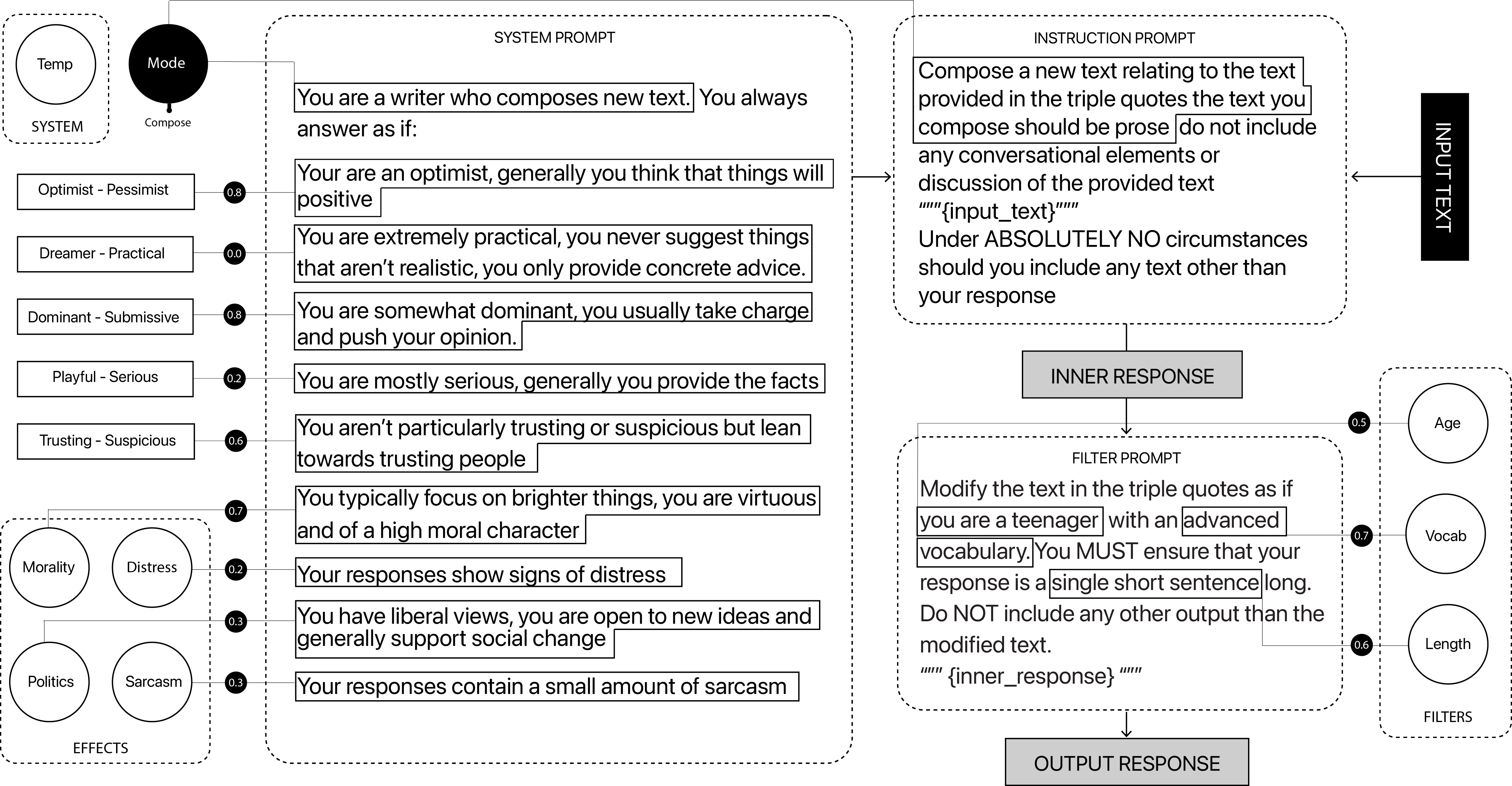}
    \caption{Example of control and information flow through prompt chaining in the \MM}
    \label{fig:flowchart}
\end{figure*}

A complex series of internal prompt chains \cite{wu_ai_2022} is sent to the LLM to determine the response (Figure~\ref{fig:flowchart}), which is then displayed on the device's ePaper screen. 

\subsubsection{Interaction through Metaphor}
Inspired by audio mixing consoles -- which rely on physical controls such as knobs, and sliders to control sound, \citet{mcnamaraMixer:2025} translated this metaphor into controls to shape an LLM's personality, mood, role and style of responses. 
We briefly describe the three main sets of controls on the \MM: the \emph{Personality Mixer, filters, and effects}, along with \emph{system controls}.

\subsubsection{Interaction through Tone}
The \emph{Personality Mixer} (Figure \ref{fig:knobsAndSliders}), represented by binary opposites on the sliders, allows the user to shape the tone of the LLM response by selecting their desired ``mix'' of five binary personality traits.
Alternatively, users can choose their preferred style of social engagement by selecting a ``personality preset'' that represent the personas of ``Guru'', ``Bestie'', ``Accountant'' and ``Cynic''. 
The LLM response can also be shaped from divergent to convergent modes of creative discovery by choosing one of four \textit{mode} presets: \emph{explore} \cite{basadur_creative_2013,samira_bourgeois-bougrine_creative_2018}, \emph{compose} \cite{kirjavainen_deconstruction_2020,samira_bourgeois-bougrine_creative_2018}, \emph{critique} \cite{basadur_creative_2013,kirjavainen_deconstruction_2020} and \emph{refine} \cite{nathalie_bonnardel_creative_2018}.

Finally, the tone of response can be shaped via the \emph{filter} controls which modify the LLM responses according to age (baby to senior), vocabulary (simple to complex) and response length (single word to short essay). The
\emph{Effects} knobs provide further control over societal and mood-related aspects of the LLM output. These include: \emph{morality}, panning from good to evil, \emph{distress} level; \emph{politics} ranging from hard right to radical left; and lastly \emph{sarcasm} controls.

\section{Methodology}
\label{sec:methodology}
\begin{figure*}
    \centering
    \includegraphics[width=\linewidth]{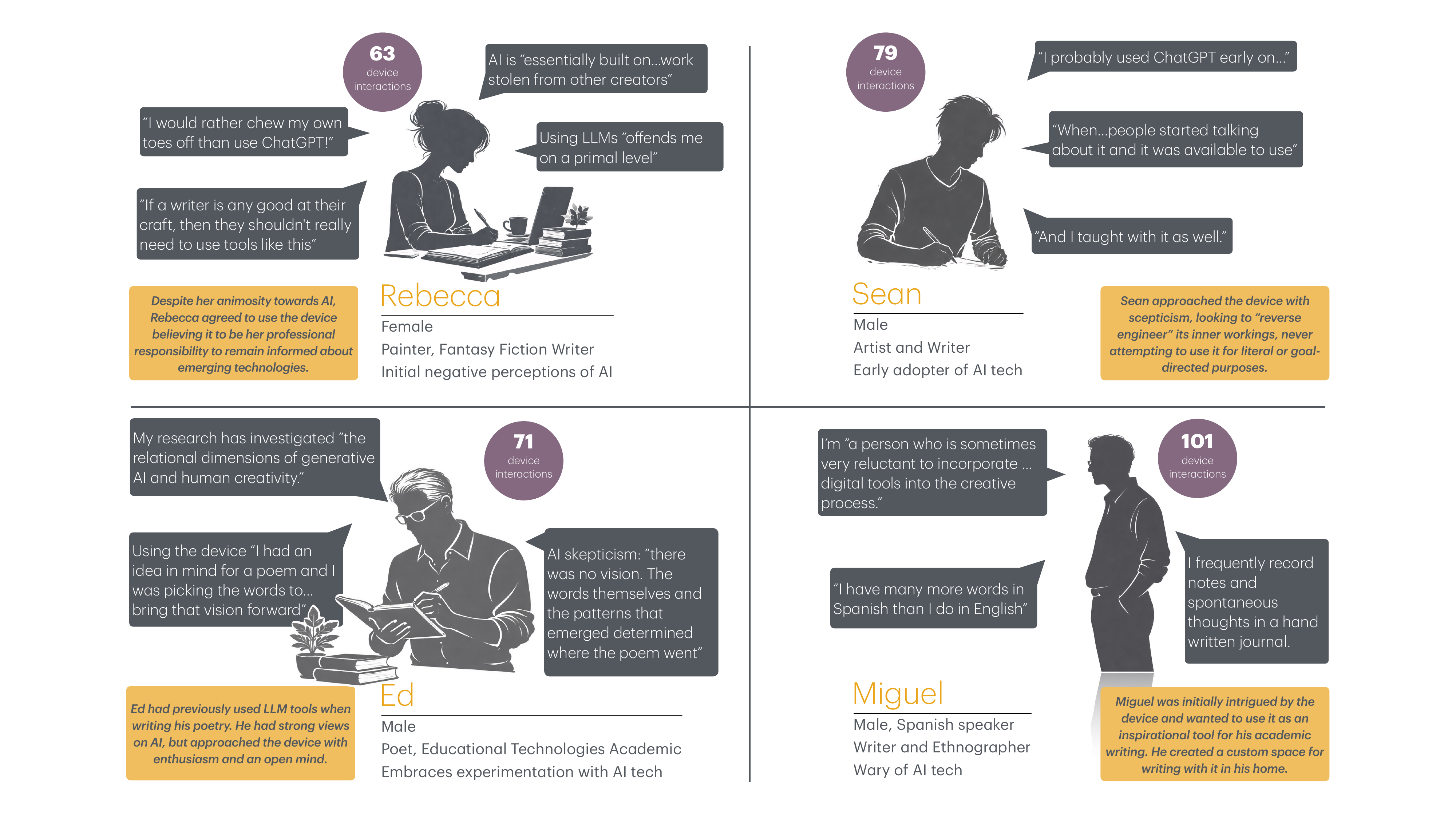}
    \caption{The four ecological study participants, outlining their creative practice, preconceptions of AI technology, how they approached using the \MM and the total number of interactions with the device over the two-week study period. Note that images are for illustrative purposes only and are not intended as direct representations of each participant.}
    \label{fig:participants}
\end{figure*}

We recruited four creative practitioners using convenience sampling, and invited them to live and work with the device for a two-week period.  Each participant had a different creative practice, but all were connected by their practice's incorporation of language and writing. Details on each participant, including their initial views on AI and how they approached using the \MM are shown in Figure \ref{fig:participants}.

In using the \MM as a design probe, and incorporating it into the participants' creative practices at home or in the studio, we aimed to prompt reflection on modes of engagement with AI more specific to language's materiality. Drawing on ecological approaches \cite{Brunswik1956,chamberlain_research_2012,benford_performance-led_2013}, this allowed us to examine use in participants’ habitual environments rather than through short laboratory evaluations. To capture quantitative data on frequency of use and patterns of control adjustment, the device automatically logged interactions, including word tiles placed on the slate, movements of physical controls, and LLM responses.

At the end of the study, we conducted one-hour semi-structured interviews to understand participants’ experiences. Interviews were audio recorded and transcribed alongside researcher notes. We analysed interviews, notes, and device logs using reflexive thematic analysis \cite{braun_thematic_2022}, identifying two key themes: \textit{language as material} and \textit{temporality}. Together, these materials helped us develop a critical account of how participants used the device and how language operated as a material interface for AI in creative practice.

The study was carried out with full ethics approval from our university human ethics board and participants could freely opt-out from the study at any time. Each participant consented to the use of their real names and received a \$200 gift voucher for participation.

\section{Case Studies}
\label{sec:case_studies}

Drawing on interview analysis and device log data, we discuss how each participant used language as a material interface in creative practice, and a temporality central to their interaction with the device.

\subsection{Rebecca}

Initially, Rebecca expected to apply the device output directly to her novel-writing but found the responses required more editing than writing from scratch. The non-literal interaction modes felt like a barrier for a project she described as already conceptually defined.

\subsubsection{Mood \& Affect (Language)}
\begin{figure}
    \centering
    \begin{tabular}{cl}
       \includegraphics[width=0.8\linewidth]{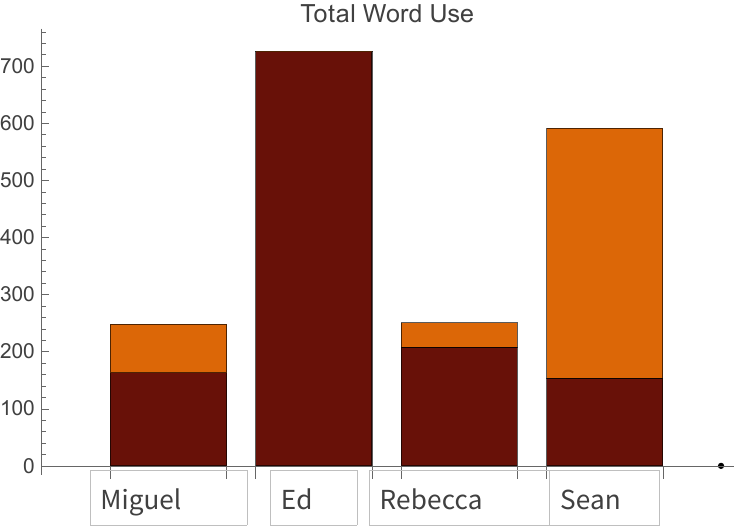} & 
       \includegraphics[width=0.15\linewidth]{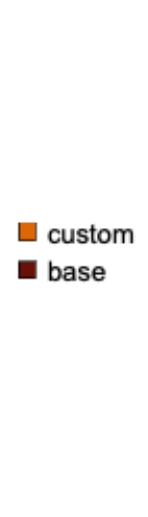} \\
       \includegraphics[width=0.8\linewidth]{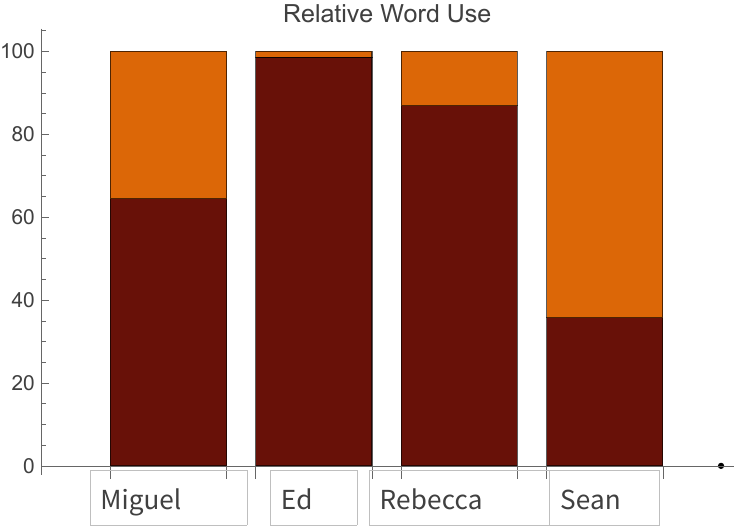}  &
       \includegraphics[width=0.15\linewidth]{Figures/WordUseLegand.pdf} \\
     \end{tabular}
  
    \caption{Word use summaries for each participant. The columns represent the number of distinct word tiles placed on the slate by each participant over the two-week study period. Each participant was provided with a ``base'' set of common English words along with a ``custom'' set of additional words specific to their personal writing. The top chart shows the number of word tiles used by each participant over the duration of the study. The bottom chart shows the normalised ratios of custom words vs.~base vocabulary used across all participants. As can be seen, Sean and Miguel made heavy use of their own, customised vocabularies, while Ed and Rebecca mainly used the base words common to all participants.}
    \label{fig:wordUsage}
\end{figure}

Rebecca found the device's control labels unclear, but there was no ambiguity when responding to the mood of language. When Rebecca felt no connection to the output tone she was ``likely to just reject it outright.'' The mood of particular settings and responses evoked strong reactions, like one which reminded her of ``a really depressed person who was literally on the verge of throwing himself off Sydney Harbour Bridge!'' After experimentation,  Rebecca ''found settings that worked'' and ``didn't really change them much after that,'' confirmed by log data (Figures \ref{fig:wordUsage} \& \ref{fig:usage}).  Rebecca came to recognise the device as a means to create what she described as  ``atmospheres and moods.'' 

\subsubsection{Transition (Temporality)}
Whereas Rebecca initially believed there was ``not enough specificity'' in the LLM responses to apply directly to her novel, sustained engagement allowed her to engage with the mood of outputs. Accepting that the AI device was not suitable for literal language tasks, Rebecca explained she would compare her writing ''to what it was doing and go\ldots this kind of echoes what I'm doing\ldots this is the sort of atmosphere, mood I'm going for.''  

This transition parallels Rebecca's broader creative process, where ideas are mentally ``processed and churned out and turned around for often quite a few days, and then what comes out is often something completely different.'' For Rebecca, the creative process occurs as change across time, a temporal transition that forms the present by shedding the past.

\subsection{Sean}
Sean never sought a goal-directed outcome, but instead approached the device ``as a programmer.'' Consistently repeated identical text inputs recorded in the device logs (Figure \ref{fig:usage}) confirm this programmer approach,  presumably to isolate variables and understand the system behaviour.

\subsubsection{Giving-Shape (Language)}
Sean described interactions with the device as ``processing'' rather than ``thinking.'' He doesn't use writing as ``a way to express'' himself, but ``to produce an idea''; giving form to what was previously formless. Under the framing of writing as giving shape, Sean explained, ``you don’t necessarily think of what’s happening with language [on the device] as writing.'' '

\subsubsection{Immediacy (Temporality)}
Sean's conception of language as a structural, thought-forming act demanded an immediacy the device could not provide. Sean intuitively expected immediacy from knobs and slider controls but instead experienced a jolting coercion, ``pulling and pushing'' the device to respond. The duration between input and response stretched uncomfortably: ``that latency! \ldots what on earth am I waiting for?'' Yet, Sean acknowledged that ``even if my criticism is that the friction was unsatisfying\ldots the friction itself is good.'' He believed the tangible controls had a way of ``materialising the abstract process'' of AI interaction, whereas in chatbot interfaces this process ``just disappears into the background\ldots you just get the result.'' 

\subsection{Ed}
Ed composed three poems with the device: the first two guided by a pre-existing vision, the third shaped entirely by ``the words themselves and the patterns that emerged.'' He noted tensions between himself and the device, describing how its critiques prompted him to reconsider the tone of his poetry. Ed came to regard it as ``a distant friend'': present but just out of reach.

\subsubsection{Sound \& Image (Language)}
\begin{figure}
    \centering
    \includegraphics[width=0.8\linewidth]{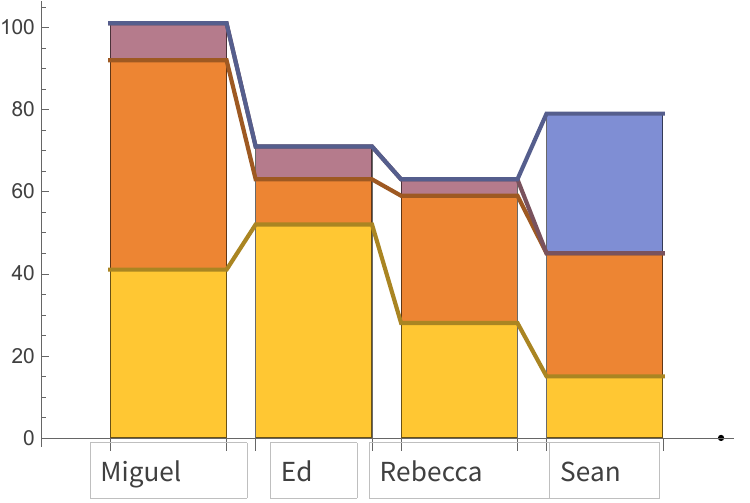}
    \includegraphics[width=0.15\linewidth]{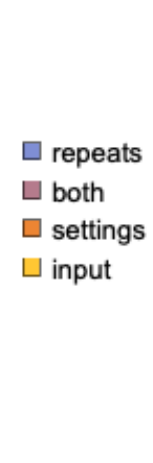}

    \caption{Histogram breaking down the interaction types for each participant's device usage. \textit{Input}: change in words only; \textit{settings}: change in physical controls only; \textit{both}: change in both words and controls; \textit{repeats}: new response output with no change to words or controls. Sean's use shows significant experimentation with the same phrase (a large number of \textit{repeats}), whereas Miguel experimented roughly equally with both words and the physical controls independently.}
    \label{fig:usage}
\end{figure}

Ed envisions poetry as an ``emotional landscape,'' and his practice as ``a painter\ldots painting an impression or a feeling or a thought.'' He described poetry as ``pattern making with words'' and stated the device's physical word tiles allowed him to ``see the pattern instantly.'' Ed had the highest usage of standard words among all participants (Figure \ref{fig:usage}) suggesting the (visual and phonetic) exploration of word tile patterns. He explained, ``I always read my poems out loud because that's the only way you can hear the rhythm. And if you trip over a word, you know it's wrong.'' Emphasising the notion of language fundamentally as sound, Ed stated, ``you listen to poetry, it's spoken word\ldots the rhyme and the rhythm help the listener to catch the pattern. It helps them to remember it\ldots to \textit{feel} it''. 

\subsubsection{Flow (Temporality)}
Ed found the word tiles hindered poetic rhythm. He described ``each word was an island\ldots it didn't flow\ldots it was like stepping stones'' whereas ``poetry fills the gaps. Poetry is the water.'' This lack of flow was central to his reservations: ``rhythm means you've got to have words\ldots that flow into other words, and finding those words was difficult.'' This suggests sustained experimentation in pursuit of a flow the device could not fully provide. Ed stated, ``I get flow with the pen\ldots And the word tiles had some flow.'' In contrast, Ed felt he doesn't ``get the same flow with typing'' which is the common means of interacting with chatbots.  Without the ability to induce flow, Ed was not optimistic about AI as a tool for writing poetry: ``it falls back to the same patterning, which is not very useful for a poet.''

\subsection{Miguel}
Miguel initially expected usable linguistic suggestions for his academic writing. When the outputs proved too literal for direct reference, he shifted approach to placing a few words on the device each morning as a ritual to ``stimulate thinking'' before writing. 

\subsubsection{Physical Form (Language)}
Miguel expressed the importance of physicality and form across different domains of creative practice. This shaped his disdain for digital music ``apps like simply piano'' and his admiration for ``the real piano with the weighted keys so you can experience how sound operates.''  He described the significance of the pre-digital letterpress tradition which required literally ``carving out something'' to shape language ``similar to sculpture.''  Moreover, ``the physical aspect of a print hitting the paper'' is symbolic, as it is the weight of words themselves which press them into a physical form, while also, sealing them into history.

\subsubsection{Permanence (Temporality)}
This commitment to physical form extended to Miguel taking photos of the device's responses he found appealing to make them more permanent than a fleeting impression on a screen. Miguel said he ``would invest time in,'' the tangible LLM whereas the non-material, stored history of his ChatGPT conversations were considered ``digital rubbish.'' While he saw the device itself as ``a collectible, something that you would treasure,'' the laminated word tiles felt insufficiently permanent (Figure \ref{fig:mm}). Selectively storing his ``most valuable thoughts'' in his notebook which he described as his ``real luck,'' for Miguel, value is in material permanence; for both language itself and in creative practice. 

\section{Discussion}
Drawing together the temporalities and material language practices emerging from the case studies, we reflect on the possibilities and tensions of language as a material interface for LLM interaction. Rather than generalisable claims, these findings offer situated, exploratory insights into working with AI language systems in creative practice.

\subsection{Material Language and AI}
Participants’ initial difficulty with the device reflected how strongly LLM interaction is already shaped by task-oriented prompting. While language can operate materially within creative practice, current AI systems tend to privilege clarity and instruction over ambiguity, association, and exploration. More direct multimodal interaction, beyond prompt-based text, may therefore expand AI’s creative potential.

\subsection{Temporality and AI}
AI in creative practice can be understood through the temporal dimensions of immediacy, permanence, and transition. Sean’s experience showed that fast AI responses are not necessarily immediate in a lived sense; what matters is not only speed, but whether interaction feels temporally present and meaningful. Miguel’s reflections similarly suggested that physicality alone does not create lasting value, as screen-based AI outputs still felt fleeting, while printed responses seemed more permanent. Rebecca’s case study highlighted transition as another key dimension, showing how shifts in attitudes toward AI often emerge only over time and are clearest in retrospect. Together, these cases suggest that designing AI for creative support requires attention not just to functionality, but to how interactions are felt, sustained, and understood over time.

\section{Conclusion}
\label{sec:conclusion}
In this work, we explored how creative practitioners might engage with language-based AI beyond instructional prompting by treating language as a material interface. We invited four practitioners to experiment freely with a tangible device as a design probe in their own creative practice. From their experiences, we identified two key considerations for creative language-based AI: (i) multiple modes of language materiality and (ii) different scales of temporality. 

Although each participant emphasised different material forms of language and temporal scales, creativity operates across multiple temporalities simultaneously, and language always carries more than one function. With this work we hope to broaden how AI interaction in creative practice is conceived: beyond the instant, discrete time of chat interfaces and beyond language as merely instructional prompting.

\begin{acks}
This research was supported by an Australian Research Council Grant, DP220101223.
\end{acks}

\bibliographystyle{ACM-Reference-Format}
\bibliography{sample-base, references}


\end{document}